\documentclass[amssymb,prl,aps,twocolumn,amsmath,showpacs]{revtex4}

\usepackage[dvips]{graphicx}

\begin{document}

\title{Magnetization-dependent $T_c$ shift in F/S/F trilayers with a strong ferromagnet}
\author{Ion C. Moraru, W. P. Pratt, Jr., Norman O. Birge}
\email{birge@pa.msu.edu}
\affiliation{Department of Physics and
Astronomy, Michigan State University, East Lansing, Michigan
48824-2320, USA}
\date{\today}

\begin{abstract}
We have measured the superconducting transition temperature $T_c$
of Ni/Nb/Ni trilayers when the magnetizations of the two outer Ni
layers are parallel (P) and antiparallel (AP).  The largest
difference in $T_c$ occurs when the Nb thickness is just above the
critical thickness at which superconductivity disappears
completely. We have observed a difference in $T_c$ between the P
and AP states as large as 41 mK - a significant increase over
earlier results in samples with higher $T_c$ and with a CuNi alloy
in place of the Ni. Our result also demonstrates that strong
elemental ferromagnets are promising candidates for future
investigations of ferromagnet/superconductor heterostructures.
\end{abstract}

 \maketitle

Heterostructures composed of ferromagnetic (F) and superconducting
(S) materials have attracted much theoretical and experimental
attention due to the rich physics produced by the interplay
between competing symmetries of the order parameters
\cite{Izyumov2002}.  In an S/F bilayer the exchange field of the
ferromagnet modulates the superconducting order parameter as it
decays inside the ferromagnet over a very short distance. Kontos
\textit{et al.} \cite{KontosPRL2001} used tunneling spectroscopy
to observe the damped oscillations of the order parameter by
measuring the density of states (DOS) for different thickness
ferromagnets. Ryazanov \textit{et al.} \cite{RyazanovPRL2001}
observed $\pi$-state Josephson coupling in an S/F/S trilayer first
by varying temperature, then later by varying the thickness of the
ferromagnet \cite{RyazanovJLTP2004}.  Earlier, several groups
\cite{ChienReich, Mercaldo1996, Muhge1996} had observed
oscillations in the critical temperature $T_c$ of S/F bilayers as
a function of the ferromagnet thickness $d_F$. Under ideal
conditions $T_c$ oscillations arise from interference between the
transmitted superconducting wave function through the S/F
interface and the wave reflected from the opposite surface of the
ferromagnet, although in some cases alternative explanations have
been proposed \cite{LazarPRB2000}. In many experiments, weakly
ferromagnetic alloys were used in order to reduce the size of the
exchange splitting in the conduction band, $E_{ex}$, and thus
increase the penetration length $\xi_F$ for Cooper pairs, where
$\xi_{F}=\hbar v_{F}/2E_{ex}$ in the clean limit and $v_F$ is the
Fermi velocity of the ferromagnet \cite{DirtyLimit}.

An alternative way to probe the influence of a ferromagnet on a
superconductor is to look for $T_c$ variations in an F/S/F
trilayer structure based on the mutual orientation of the two
ferromagnet magnetizations \cite{BuzdinEPL1999,TagirovPRL1999}.
This effect was observed \cite{GuPRL2002} and later reproduced
\cite{PotenzaPRB2005} in a Cu$_{1-x}$Ni$_x$/Nb/Cu$_{1-x}$Ni$_x$
system, where a weak ferromagnet was used because it is "less
devastating to superconductivity."  The largest difference in
$T_c$ observed between the antiparallel (AP) and parallel (P)
states of the F-layer mutual magnetizations was only 6 mK when
$T_c$ was 2.8 K.  Unlike other experiments
\cite{KontosPRL2001,RyazanovPRL2001,RyazanovJLTP2004,ChienReich,
Mercaldo1996, Muhge1996} that require the ferromagnet thickness to
be comparable to $\xi_F$, however, a positive feature of this
experiment is that the $T_c$ difference is predicted to persist
even for thick F layers \cite{TagirovPRL1999,BuzdinEPL1999}. Thus
it proves advantageous in studying systems with strong elemental
ferromagnets, which have extremely short values of $\xi_F$.

Experimental studies of F/S systems with strong ferromagnets are
of interest because they provide new challenges to theory, which
does not yet address the full complexity of the ferromagnetic
state with its different DOS and $v_F$ of the majority and
minority spin bands.  Furthermore, pure elemental ferromagnets are
in the clean limit, $\xi_F < l_F$ where $l_F$ is the mean free
path; this complicates use of the the popular Usadel equations
normally applied to the dirty limit. We are motivated to work in
the limit of thick ferromagnetic layers, in anticipation of future
situations where superconducting order with spin-triplet symmetry
is induced in a superconductor surrounded by ferromagnets with
noncollinear magnetizations \cite{Bergeret2004}. (When
$d_F\gg\xi_F$, the singlet component of the order parameter is
completely damped.)  Lastly, we wish to understand whether, in an
F/S/F system with a strong ferromagnet, a large difference in
$T_c$ between the P and AP states can be achieved, as envisioned
in the proposals for a superconducting spin switch
\cite{BuzdinEPL1999,TagirovPRL1999}. The Ni/Nb system has been
shown to be a viable candidate for experiments on F/S systems
\cite{BlumPRL2002,Sidorenko2003}.  In this paper we will show that
Ni/Nb/Ni trilayers exhibit a significant $T_c$ shift depending on
the mutual orientation of the magnetizations of the two Ni layers.
\begin{figure}[ptbh]
\begin{center}
\includegraphics[width=3.2in]{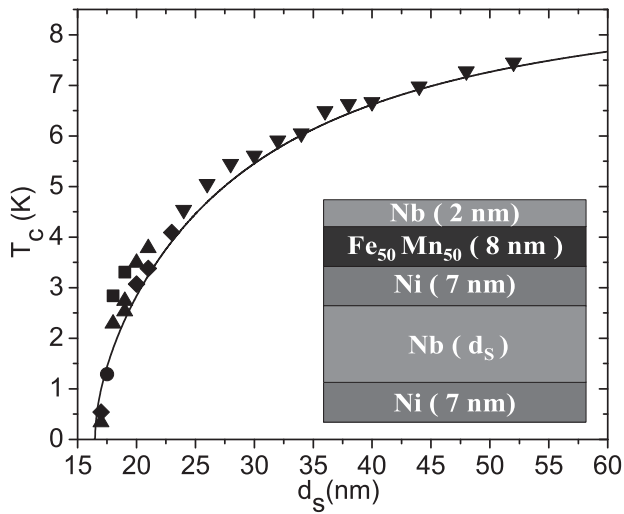}
\end{center}
\caption{Critical temperature vs. Nb thickness for
Ni(7)/Nb($d_s$)/Ni(7)/Fe$_{50}$Mn$_{50}$(8)/Nb(2) samples (all
thicknesses are in nm). Different symbols represent different
sputtering runs. The solid line represents the theoretical fit.
Inset: Schematic cross-section of the samples.} \label{Fig1}
\end{figure}

Sets of Ni(7)/Nb($d_{s}$)/Ni(7)/Fe$_{50}$Mn$_{50}$(8)/Nb(2)
multilayers (all thicknesses are in nm) were directly deposited
onto Si substrates by magnetically-enhanced triode dc sputtering
in a high vacuum chamber with a base pressure in the low $10^{-8}$
Torr and an Ar pressure of $2.0\cdot10^{-3}$ Torr.  The Ni
thickness of 7 nm was chosen to be much longer than $\xi_F$, which
we estimate to be 0.8 nm using $2E_{ex}$ = 0.23 eV and $v_F =
0.28\cdot10^{6}$ m/s for the majority band
\cite{PetrovykhAPL1998}.  The purpose of the FeMn is to pin the
magnetization direction of the top Ni layer by exchange bias
\cite{NoguesJMMM1999}. The non-superconducting Nb capping layer
protects the FeMn from oxidation. After deposition, the samples
were heated to $180^{\circ}$C under vacuum, just above the
blocking temperature of FeMn, and cooled in an applied field of
200 Oe in the plane of the multilayer. This procedure pins the top
Ni layer while leaving the bottom Ni layer free to rotate in a
small applied magnetic field.

Four-probe resistance measurements with the current in the plane of
the multilayer were performed to determine $T_c$.  Samples had
lateral dimensions 4.3 mm x 1.6 mm.  The $T_c$ of each sample was
defined to be the temperature at which the resistance dropped to
half its normal state value. Fig. 1 shows the results for $T_c$
measurements for samples from several sputtering runs, where $d_s$
was varied between 16-52 nm. $T_c$ shows a strong dependence on the
superconductor thickness close to a critical thickness, $d_s^{cr}$,
where the sensitivity to ferromagnetism is enhanced. There is no
superconductivity above 36 mK for $d_s < d_s^{cr} \approx 16.5$ nm.
\begin{figure}[ptbh]
\begin{center}
\includegraphics[width=3.2in]{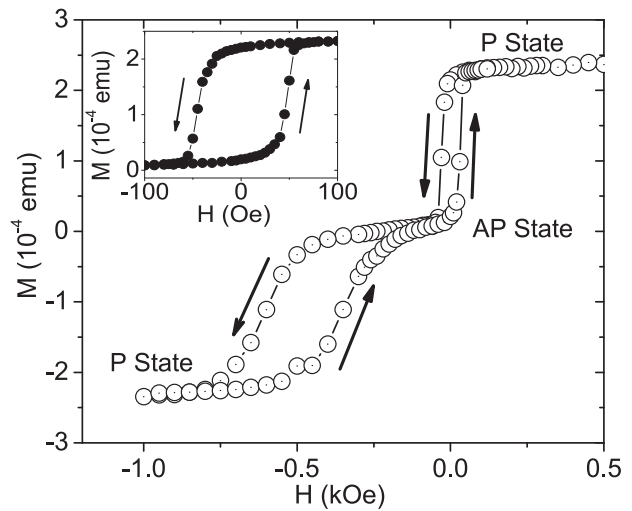}
\end{center}
\caption{Magnetization vs. applied field for a $d_s = 18$ nm
sample measured at $T = 100$ K. At $\pm$50 Oe the free bottom Ni
layer switches while the pinned top Ni layer switches at -500 Oe.
Inset: minor loop measured at $T = 2.29$ K showing the switching
of the free Ni layer.} \label{Fig2}
\end{figure}

The magnetic configuration of our structures was verified on
simultaneously sputtered samples of larger lateral size, in a
SQUID magnetometer. Fig. 2 shows a plot of magnetization vs.
applied field $H$ for a sample with $d_s$ = 18 nm taken at 100 K.
The narrow hysteresis loop near $H$ = 0 shows the switching
behavior of the free Ni layer with a coercive field $H_c$ = 35 Oe.
The wider loop shows switching of the pinned layer and is shifted
to nonzero $H$ due to the exchange bias between the top Ni layer
and the FeMn. Applied fields of $\pm$100 Oe switch the spin-valve
between the P and AP configurations.   The nearly zero net
magnetization observed at -100 Oe indicates very good AP alignment
between the pinned and free Ni layers, while the nearly saturated
magnetization observed at +100 Oe indicates good P alignment.
Similarly good alignment of the P and AP states can be achieved at
low temperature.  The inset to Fig. 2 shows a minor hysteresis
loop with $H_c \approx$ 50 Oe taken at 2.29 K, which corresponds
to the middle of the superconducting transition for this sample.
We obtain the same behavior for temperatures above and below the
transition temperature.

Measurements of $T_c^P$ and $T_c^{AP}$ were performed by
alternating the applied field between +100 and -100 Oe, as the
temperature was slowly decreased through the transition region.
The largest shift in critical temperature, $\Delta T_c \equiv
T_c^{AP} - T_c^P$, should occur in samples with the Nb thickness
close to $d_s^{cr}$. Fig. 3 shows a plot of $R$ vs. $T$ for a
sample with $d_s$ = 17 nm, measured in a dilution refrigerator.
Two distinct transitions are observed for P and AP alignment, with
a separation in temperature $\Delta T_c \approx 28$ mK.  A second
sample with $d_s$ = 17 nm showed a $\Delta T_c \approx 41$ mK, but
with a slightly broader transition centered at 0.34 K. Samples
with $d_s$ = 18 nm and $T_{c}$ between 2 and 3 K exhibit values of
$\Delta T_c$ of only a few mK, similar to the CuNi/Nb/CuNi samples
measured previously \cite{GuPRL2002,PotenzaPRB2005}.

\begin{figure}[ptbh]
\begin{center}
\includegraphics[width=3.2in]{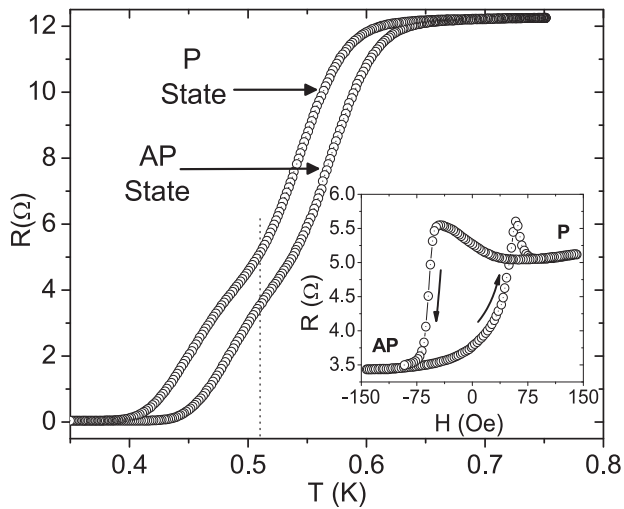}
\end{center}
\caption{Resistance vs. temperature for the P and AP states of a
$d_s = 17$ nm sample measured in $\pm$100 Oe.  Two distinct
transitions are observed, with $\Delta T_c =$ 28 mK. Inset:
Resistance vs. applied field at $T = 0.51$ K (dotted line in main
graph).} \label{Fig3}
\end{figure}

The inset to Fig. 3 shows a plot of $R$ vs. $H$ for the first
$d_s$ = 17 nm sample at a temperature in the middle of the
transition (0.51 K).  The data clearly show well-established P and
AP states at $\pm 100$ Oe, respectively, with a difference in
resistance of 1.5 $\bf{\Omega}$.  Above the transition the
resistance does not change perceptibly when switching from P to AP
alignment.  An interesting feature of the $R$ vs. $H$ curve is the
behavior of the resistance as the field is swept down from +150 Oe
towards -50 Oe and as the field is swept up from -150 Oe towards
+50 Oe.   In both cases the resistance increases to a value higher
than that of the P state after the field passes through zero. We
believe this behavior involves the breaking of the free
ferromagnetic layer into domains when $H \approx H_c$. The
domain-wall fringe fields penetrate the superconductor, thus
suppressing $T_c$ slightly and producing a higher resistance. Note
that this effect is \textit{opposite} to that observed by other
groups \cite{RusanovPRL2004,KinseyIEEE2001}, where inhomogeneous
magnetization led to enhanced superconductivity in F/S bilayers.
In those experiments, the domain size must be smaller than the
superconducting coherence length so that the Cooper pairs sample
multiple domains \cite{ChampelPRB2005}, and the magnetic field
penetrating into the superconductor must be small.

The critical temperature of F/S/F trilayers in the P and AP states
has been calculated theoretically by several groups
\cite{TagirovPRL1999,BuzdinEPL1999,YouPRB2004,FominovJETP2003,BaladiePRB2003}.
Since many experiments employ ferromagnetic alloys, the usual
approach involves solving the Usadel equations in the dirty limit
for both the superconductor and the ferromagnet.  (The dirty limit
applies to S when $l_S < \xi_{BCS} = \hbar v_S \gamma/\pi^2
k_{B}T_{c0}$, and to F when $l_F < \xi_F$, where $l_S$ and $l_F$
are the electron mean free paths in S and F, and $T_{c0}$ is the
transition temperature of the bulk superconductor.)  In our case,
however, the ferromagnetic metal is both pure and strong, thus in
the clean limit $l_F > \xi_F$.  Hence we use the theory of
\cite{TagirovPRL1999} as modified in section 3.2 of
\cite{TagirovPhysicaC1998} to make it more appropriate for the
clean limit. This theory does not, however, incorporate a full
description of the majority and minority spin bands of a strong
ferromagnet, with different DOS, $v_F$, and transmission
coefficients. The expression for the normalized critical
temperature of the trilayer is
\begin{equation}
\textrm{ln}\,t_c + Re\,\Psi\left(\frac{1}{2} +
\frac{2\phi^2}{t_c(d_s/\xi_S)^2}\right)-\Psi\left(\frac{1}{2}\right)=0,
\end{equation}
where $t_c \equiv T_c/T_{c0}$ and $T_{c0}$ is the critical
temperature of an isolated Nb film of the same thickness as the
one in the trilayer.  The function $\phi$ is determined from the
condition $\phi\, \textrm{tan}\phi = R$ for the P state or
$(\phi\, \textrm{tan}\phi -
R')(R'\,\textrm{tan}\phi+\phi)-(R'')^2\textrm{tan}\phi=0$ for the
AP state, where the complex function $R=R'+iR''$ is given by:
\begin{equation}
R=\frac{d_s}{\xi_S}\frac{N_F v_F \xi_S}{2N_S
D_S}\frac{1}{\sqrt{1-i\xi_F/l_F}+2/T_F}
\end{equation}
Eq. (2) is valid when the ferromagnets are thick enough so that
the tanh functions in \cite{TagirovPhysicaC1998} can be set to 1.
This assumption is validated by data on Nb/Ni bilayers
\cite{Sidorenko2003} where oscillations in $T_c(d_F)$ are
completely damped for $d_F > 4$ nm. The dimensionless parameters
that enter into this theory are the ratios $d_s/\xi_S$,
$\xi_F/l_F$, the S/F interface transparency $T_F$, and the
combination $N_F v_F \xi_S/2 N_S D_S$. $N_F$ and $N_S$ are the
densities of states at the Fermi energy of the F and S layers,
$v_F$ is the Fermi velocity of the ferromagnet, and $D_S$ is the
diffusion constant of the superconductor.

To avoid fitting the data with four free parameters, we follow the
strategy outlined by Lazar \textit{et al.} \cite{LazarPRB2000} and
by Sidorenko \textit{et al.} \cite{Sidorenko2003}.  We determine
the superconducting coherence length, $\xi_S$, from measurements
of the critical field vs. temperature of isolated Nb films, with
the magnetic field applied perpendicular to the film plane. For
films in the thickness range 20-50 nm, the values of $\xi_S$ are
close to 6 nm, which we use for our fits
\cite{NoteDiffusionConstant}.  From the asymptotic form of Eq. (1)
as $t_c \rightarrow 0$, one finds $2\phi^2/(d_s^{cr}/\xi_S)^2 =
1/4\gamma$, where $\gamma = 1.781$. Substituting $d_s^{cr}\approx
16.5$ nm and using Eq. (2) (while ignoring the small imaginary
term), we obtain the constraint
\begin {equation}
\frac{N_F v_F \xi_S}{2 N_S D_S(d_s^{cr})} \frac{1}{1+2/T_F}
\approx \frac{\phi^{cr}\textrm{tan}\phi^{cr}}{(d_s^{cr}/\xi_s)}=
0.24
\end{equation}
Estimates of the product $N_F v_F$ vary substantially in the
literature. From \cite{NiDOS} and \cite{PetrovykhAPL1998}, we
obtain respectively $N_F=1.77\cdot10^{48}$ J$^{-1}$m$^{-3}$ and
$v_F = 0.28\cdot10^{6}$ m/s.  Fierz \textit{et al.}
\cite{Fierz1990}, however, quote $\rho_F\, l_F = 0.7-2.3$
f$\Omega$m$^2$ for Ni, which when combined with the Einstein
relation $1/\rho_F\,l_F = N_F v_F e^2/3$ imply values 3-10 times
smaller for $N_F v_F$. Combining these values with
$N_S=5.31\cdot10^{47}$ J$^{-1}$m$^{-3}$ \cite{NbDOS} and using our
measured $D_S(d_s^{cr}) = 2.8\,\cdot\,10^{-4}$ m$^2$/s, we obtain
$T_F = 0.05-0.6$.  The bulk resistivity of our sputtered Ni films
at 4.2 K is $\rho_F = 33$ n$\Omega$m, which leads to values of
$l_F$ between 7 and 70 nm, given the range in $\rho_F\,l_F$ quoted
above. Since the Ni used in our trilayers is thin, $l_F$ is
probably limited by surface scattering, so we use the lower
estimate $l_F = 7$ nm, hence $\xi_F/l_F \approx$ 0.1. In fact, the
fit to $T_c(d_s)$ is quite insensitive to the values of $T_F$ and
$\xi_F/l_F$.  We used $\xi_F/l_F = 0.1$ and $T_F = 0.3$ to obtain
the curve shown in Fig. 1, which fits the data remarkably well.
\begin{figure}[ptbh]
\begin{center}
\includegraphics[width=3.2in]{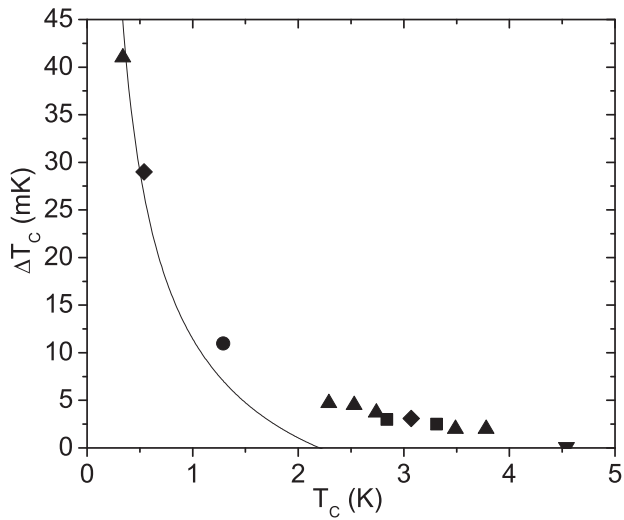}
\end{center}
\caption{Symbols: $\Delta T_c$ vs. $T_c$ for our 11 thinnest
samples. The line represents a fit using $\xi_F/l_F = 0.7$ and
$T_F = 1.0$, values larger than our best estimates.} \label{Fig4}
\end{figure}

A more stringent test of the theory is the prediction of $\Delta
T_c$, which depends sensitively on both $T_F$ and $\xi_F/l_F$.
Thickness deviations from nominal values produce scatter in plots of
$T_c$ or $\Delta T_c$ vs. $d_s$, therefore Fig. 4 shows a plot of
$\Delta T_c$ versus $T_c$.  If we calculate $\Delta T_c$ using our
best estimate of $\xi_F/l_F$ and the upper limit of $T_F$ given
above, the maximum value of $\Delta T_c$ is only a few mK when $T_c$
is well below 1 K -- hardly visible on Fig. 4.  If we relax the
constraints we have placed on the parameters, and instead try to
produce the best fit to the $\Delta T_c(d_s)$ data, we find that a
reasonable fit can be obtained when $\xi_F/l_F$ is allowed to be
much larger than our original estimate.  Fig. 4 shows a fit using
$\xi_F/l_F = 0.7$ and $T_F = 1.0$.  Similar curves can be produced
by simultaneously varying $\xi_F/l_F$ and $T_F$ while keeping their
product nearly constant. Fitting the $\Delta T_c$ data requires
letting $\xi_F/l_F$ exceed our estimate substantially.  Our $l_F$
estimate may be too large, because the resistivity is dominated by
the longer of the majority or minority band $l_F$, whereas the F/S
proximity effect depends on the shorter of the two
\cite{LazarPRB2000}. A shorter $l_F$ is also implied by the
observation of complete damping of $T_c$ oscillations in Nb/Ni
bilayers for $d_F > 4$ nm \cite{Sidorenko2003}.  Nevertheless,
producing a reasonable fit to our $\Delta T_c$ data entails either
increasing $\xi_F/l_F$ beyond the clean limit, or increasing $T_F$
beyond our original estimate.

In conclusion, we have observed a large difference in $T_c$
between the P and AP magnetic states of Ni/Nb/Ni trilayers, with
$T_c^P < T_c^{AP}$. Recently, Ruzanov \textit{et al.}
\cite{RusanovCondMat2005} reported a $T_c$ difference between the
P and AP states of a Ni$_{0.8}$Fe$_{0.2}$/Nb/Ni$_{0.8}$Fe$_{0.2}$
trilayer, but with $T_c^P > T_c^{AP}$.  Understanding these
opposing behaviors in F/S systems with strong ferromagnets will
require further experiments, as well as theoretical models able to
account for the complexity of real ferromagnets
\cite{VodopyanovJETP20003}.

We are grateful to R. Loloee and J. Bass for fruitful discussions.
This work was supported by NSF grants DMR 9809688, 0405238, 0202476
and by the Keck Microfabrication Facility.


\begin{thebibliography} {99}
\bibitem{DirtyLimit} The experiments in refs. [2-4] were in the dirty limit,
so $\xi_{F}=\sqrt{\hbar D_{F}/2E_{ex}}$, where $D_{F}$ is the
diffusion constant.
\bibitem{NoteDiffusionConstant}  From $\xi_S = \sqrt{\hbar D_S/2\pi k_B T_{c0}}$, we obtain $\xi_S =$ 7-8 nm for
$d_s =$ 20-50 nm, where $D_S$ is determined from resistivity
measurements and $T_{c0}$ is the critical temperature of isolated
Nb films.  We find $D_S$ (m$^2$/s) $=
1.8\cdot10^{-4}+5.9\cdot10^{-6} d_s$ and $T_{c0}\textrm{(K)}$ =
9.1 - 43/$d_s$, with $d_s$ in nm.
\end{thebibliography}
\end{document}